\documentclass[aps,prl,twocolumn,groupedaddress]{revtex4}
\usepackage{graphicx}

\begin{document}


\title{Effect of the shape anisotropy on the magnetic configuration of (Ga,Mn)As \\and its evolution with temperature}



\author{K. Hamaya\footnote{E-mail address: hamaya@iis.u-tokyo.ac.jp
}}
\affiliation{%
Institute of Industrial Science, The University of Tokyo, 4-6-1 Komaba, Meguro-ku, Tokyo 153-8505, Japan.
}%
\author{T. Taniyama}
\affiliation{%
Materials and Structures Laboratory, Tokyo Institute of Technology, 4259 Nagatsuta, Midori-ku, Yokohama 226-8503, Japan. 
}%
\author{T. Koike, and Y. Yamazaki}
\affiliation{%
Department of Innovative and Engineered Materials, Tokyo Institute of
Technology, 4259 Nagatsuta, Midori-ku, Yokohama 226-8502, Japan.
}%



\date{\today}

\begin{abstract}
We study the effect of the shape anisotropy on the magnetic domain configurations of a ferromagnetic semiconductor (Ga,Mn)As/GaAs(001) epitaxial wire as a function of temperature. Using magnetoresistance measurements, we deduce the magnetic configurations and estimate the relative strength of the shape anisotropy compared with the intrinsic anisotropies. Since the intrinsic anisotropy is found to show a stronger temperature dependence than the shape anisotropy, the effect of the shape anisotropy on the magnetic domain configuration is relatively enhanced with increasing temperature. This information about the shape anisotropy provides a practical means  of designing nanostructured spin electronic devices using (Ga,Mn)As.

\end{abstract}


\maketitle

\section{INTRODUCTION}

Patterned ferromagnetic semiconductor (Ga,Mn)As has shown several fascinating magnetic properties on the sub-micrometer or nanometer length scale.\cite{Ruster,Eid} Our previous work on the magnetoresistance clearly showed that the magnetic domain configurations of micro-patterned (Ga,Mn)As wires can be controlled by changing the shape anisotropy through modification of the hole concentration,\cite{Hamaya3} and therefore the magnetic shape anisotropy was found to play a crucial role in the formation process of the magnetic domain configurations.  In general views, it is believed that static magnetic energetics provide a qualitative explanation for such magnetic domain configuration, based on the total magnetostatic energy expressed as\cite{Hamaya3} 
\begin{eqnarray}
E & = &K_{u}sin^{2} (\varphi-45^\circ)+(K_{c}/4)sin^{2} 2\varphi + K_{s}sin^{2} \varphi \nonumber \\
   & &-MHcos (\varphi-\theta), 
\end{eqnarray}
where {\it K$_{u}$}, {\it K$_{c}$} and {\it K$_{s}$} are the in-plane [110] uniaxial, $\left\langle 100 \right\rangle$ cubic, and shape anisotropy 
constants, respectively, {\it M} is the magnetization, {\it H} is the applied field strength. $\varphi$ and $\theta$ are the magnetization and applied field direction with respect to [100], respectively. \cite{Hamaya3} In fact, numerical calculation using this formalism is in good agreement with the experimental data. 

However, if (Ga,Mn)As is used as a potential material for spin electronic devices in a temperature variable environment, the formalism should be amended so as to include temperature variation in the anisotropy constants because the intrinsic magnetic anisotropy of (Ga,Mn)As significantly changes with temperature:\cite{Welp,Hamaya2,Sawicki} the temperature dependence of the magnetic anisotropy was found to have a great effect on the domain configurations and magnetization reversal processes in  (Ga,Mn)As.\cite{Welp,Hamaya2} Therefore, of most essence  in designing and manipulating the magnetic configuration is to understand the relative relationship between the intrinsic anisotropy and the shape anisotropy as a function of temperature.

In this paper, we examine the effect of the shape anisotropy on the magnetic domain configuration as a function of temperature in a micro-patterned (Ga,Mn)As epilayer. Magnetotransport measurements and numerical calculation reveal that the contribution of the shape anisotropy is enhanced with increasing temperature opposed to a reduction in the magnetization with temperature. This behavior can be understood as a steep decrease in the relative contribution of $\left\langle 100 \right\rangle$ cubic magnetocrystalline anisotropy which is intrinsic to zinc-blende (Ga,Mn)As. We demonstrate that competition between the $\left\langle 100 \right\rangle$ cubic magnetocrystalline anisotropy and the shape anisotropy along [100] occurs far below the Curie temperature of (Ga,Mn)As.
\begin{figure}
\includegraphics[width=8cm]{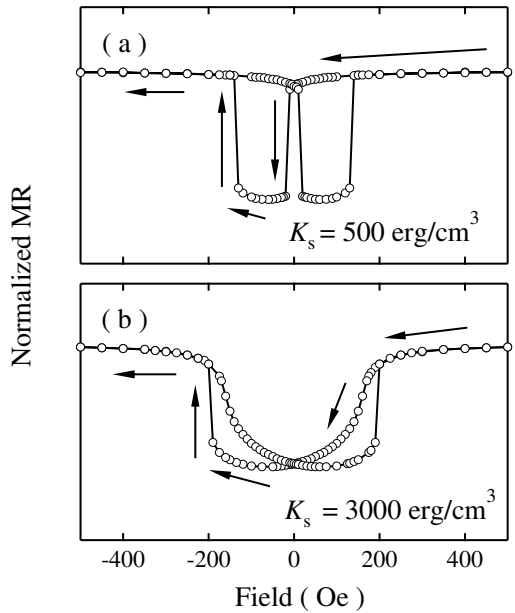}
\caption{Normalized MR curves obtained by numerical calculation using Eq. (1). ((a) {\it $K_{s}$}$ =$ 500 erg/cm$^{3}$ and (b) {\it $K_{s}$}$ =$ 3000 erg/cm$^{3}$).}
\end{figure}

\section{Experimental}
In-plane magnetized Ga$_{0.962}$Mn$_{0.038}$As film with a thickness of 100 nm was grown on a semi-insulating GaAs (001) substrate using molecular beam epitaxy (MBE). Prior to low temperature growth of (Ga,Mn)As at 235$^{\circ}$C, 
a 400 nm-thick GaAs buffer layer was grown at 590$^{\circ}$C. The hole carrier concentration $p$ was measured to be $\sim$3.5 $\times$ 10$^{20}$ cm$^{-3}$ by an electrical capacitance-voltage method at room temperature.
The magnetic properties were examined using a superconducting quantum interference device (SQUID) magnetometer. The temperature dependent magnetization curve shows that the magnetization disappears at around 60 K, i.e., $T$$_{c}$ $\sim$ 60 K. The film was patterned into a  1.5 $\mu$m-wide $\times$ 200 $\mu$m-long wire structure along GaAs [100] with ohmic contacts of Ti/Au. A scanning electron micrograph of a typical sample was shown in our previous paper.\cite{Hamaya4} Longitudinal magnetotransport measurements were performed using a standard four-point ac method at various temperatures. A magnetic field was applied parallel to the film plane and samples can be rotated in the film plane to change an angle $\theta$.
\begin{figure}
\includegraphics[width=8.5cm]{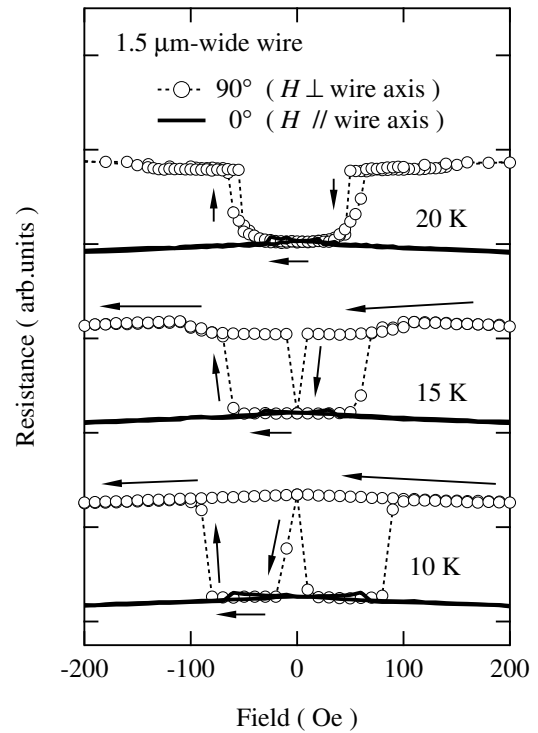}
\caption{Temperature dependence of MR curves for a wire sample with a width of 1.5 $\mu$m in $H // I$ and $H \bot I$. }
\end{figure}

\section{Results and Discussion}
Before presenting experimental data on the magnetotransport of (Ga,Mn)As wires, it is worth examining the longitudinal magnetotransport feature assuming a coherent rotation model of a single domain magnetic element.\cite{Tang,Goen} Since we can determine the magnetization direction $\varphi$ for $\theta$ $=$ 90$^{\circ}$ based on the conditions $\partial${\it E}/$\partial$$\varphi$ $=$ 0 and $\partial$$^2${\it E}/$\partial$$\varphi$$^2$ $>$ 0 for local minima of the total energy in Eq.(1),\cite{Hamaya3} the corresponding magnetoresistance (MR) curve is reproduced using {\it R} $\propto$ cos$^2$$\varphi$.  Figure 1 shows MR curves for different {\it $K_{s}$} values at $\theta$ $=$ 90$^{\circ}$, calculated using {\it $K_{c}$} $=$ 3500 erg/cm$^{3}$, {\it $K_{u}$} $=$ 1800 erg/cm$^{3}$, and $M =$ 27.5 emu/cm$^{3}$.\cite{Hamaya3} Sweeping magnetic field from positive to negative, we observe an abrupt MR jump at around $H =$ -20 Oe for {\it $K_{s}$}$ =$ 500 erg/cm$^{3}$ (Fig. 1(a)). A small change in the MR associated with magnetization rotation is also seen in the field range $H =$ -20 Oe to -130 Oe, clearly showing that the magnetization is stable along the wire axis. On the other hand, a gradual decrease in the MR is seen in magnetic fields from 500 Oe to 0 Oe for {\it $K_{s}$} $=$ 3000 erg/cm$^{3}$ (Fig. 1(b)), similar to MR curves associated with magnetization rotation from perpendicular to parallel to the wire axis. This variation in the MR curves is clearly related to the effect of the shape anisotropy on the magnetization reversal, in good agreement with our previous experimental demonstration;\cite{Hamaya3} the MR features of (Ga,Mn)As are varied with wire width as well as hole concentration.

We hereafter show experimental data on the magnetization reversal processes of a wire sample with a width of 1.5 $\mu$m as a function of temperature. Figure 2 depicts the MR curves of the wire at various temperatures in magnetic fields perpendicular ($H \bot I$) or parallel ($H // I$) to the wire axis. The MR curves at 10 K show a very small hysteretic feature and no steep MR jumps for $H // I$, while a clear hysteretic behavior is observed for $H \bot I$. According to the anisotropic magnetoresistance (AMR) of (Ga,Mn)As, the MR shows a maximum when the magnetization direction is perpendicular to current flow (wire axis),\cite{Hamaya2} indicating that the magnetization direction remains along the wire axis for $H // I$, whereas the magnetization switches between parallel and perpendicular to the wire axis for $H \bot I$. With increasing temperature from 10 K to 20 K, the MR features change strikingly as shown in Fig. 2. The remanent magnetization direction tends to align along the wire axis even for $H \bot I$ at 15 K and consequently the remanent magnetization for both $H // I$ and $H \bot I$ aligns along the wire axis at 20 K. These results indicate that the effect of the shape anisotropy is likely to overcome the intrinsic magnetic anisotropies within this very narrow temperature region and the magnetic configuration is potentially  controlled through a slight change in the external environment. Also, we note that the effect of the shape anisotropy on the magnetic domain configuration is becoming pronounced in spite of a reduction in the magnetization, which generally decreases shape anisotropy.
\begin{figure}
\includegraphics[width=8.5cm]{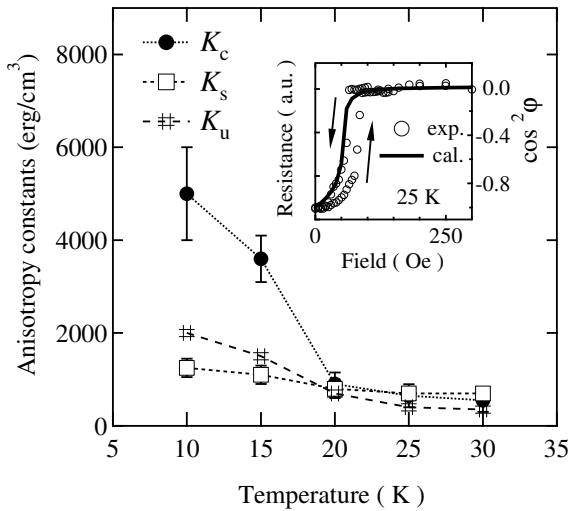}
\caption{Anisotropy constants as a function of temperature, estimated from fitting Eq. (1) to the experimental data. The inset shows a representative MR curve measured at 25 K (open circles) and the corresponding fitting curve (solid line). }
\end{figure}

In order to explore a possible cause of the characteristic variation in the MR described above, we estimate the anisotropy constants ({\it K$_{u}$}, {\it K$_{c}$} and {\it K$_{s}$}) at various temperatures from a fit of Eq. (1) to the experimental data. Provided that coherent rotation is valid for $H \bot I$, the MR curves are reproduced under the conditions $\partial${\it E}/$\partial$$\varphi$ $=$ 0 and $\partial$$^2${\it E}/$\partial$$\varphi$$^2$ $>$ 0, as stated before.\cite{Hamaya3} The fitting is done in the demagnetizing process from $+$ 500 Oe to 0 Oe. The inset of Fig. 3 displays an experimental MR hysteresis curve (open circles) measured at 25 K together with a calculation curve (solid line) using Eq. (1), for example. For the fitting, we use magnetization values obtained from SQUID measurements at various temperatures.  The anisotropy constants {\it K$_{u}$}, {\it K$_{c}$} and {\it K$_{s}$} are plotted in Fig. 3 as a function of temperature. It should be noted that the reduction in {\it K$_{c}$} is more significant than that in {\it K$_{u}$} and {\it K$_{s}$}.\cite{Welp2,Liu} Also, the value of {\it K$_{c}$} is comparable to that of {\it K$_{s}$} at 20 K, indicating that the relative contribution of {\it K$_{s}$} is enhanced remarkably above $\sim$20 K in spite of the reduced absolute value of {\it K$_{s}$} with increasing temperature. Further, we note that the value of {\it K$_{s}$} at 10 K is compatible with a value of $\sim$2000 erg/cm$^{3}$ calculated using the equation, {\it K$_{s}$} $=$ $\frac{M^{2}}{4\mu_{0}}$, which is valid for an infinitely long ferromagnetic wire, where $\mu_{0}$ is the magnetic permeability of vacuum. 
\begin{figure}
\includegraphics[width=8.5cm]{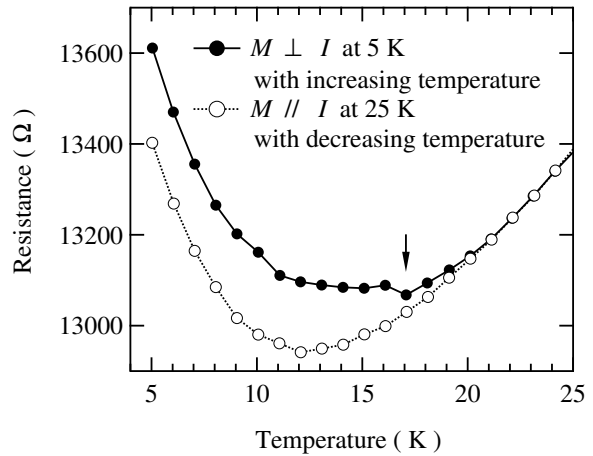}
\caption{Temperature dependent resistance of a wire sample with a width of 1.5 $\mu$m for $M // I$ and $M \bot I$. }
\end{figure}

We also deduce the temperature variation in the magnetic configuration at zero field. Figure 4 shows that the temperature dependent resistance recorded with decreasing temperature after the saturation of the magnetization in 10 kOe at 25 K (open circles) and that recorded with increasing temperature after the saturation in 10 kOe at 5 K (closed circles). It is seen that the curve shown by the closed circles merges with the curve shown by the open circles at around 20 K, accompanied with an abrupt reduction in the resistance near 17 K, indicating that the contribution of {\it K$_{s}$} along [100] becomes significant at 17 K. This behavior also implies that the contribution of {\it K$_{c}$} is significantly reduced far below the $T$$_{c}$ of 60 K. From these results, the shape anisotropy plays a  predominant role in the formation of magnetic domain configurations above 17 K, although the origin of the rapid decrease in the {\it K$_{c}$} with increasing temperature is still unclear: recent studies also have discussed a rapid decrease in the {\it K$_{c}$} for in-plane magnetized (Ga,Mn)As.\cite{Liu,Welp,Welp2,Sawicki,Hamaya2} To understand more details of the magnetic anisotropy for in-plane magnetized (Ga,Mn)As epilayers,  further study is required taking microscopic magnetic information including magnetic inhomogeneity as suggested in our previous report \cite{Hamaya} into account. 

\section{Summary}
We have examined temperature variation in the shape anisotropy and its effect on the magnetic configuration of a patterned (Ga,Mn)As structure using magnetoresistance measurements. The relative contribution of the shape anisotropy is found to be significantly enhanced with increasing temperature due to a reduction in the cubic magnetic anisotropy far below the Curie temperature. The information shown here is the first clear manifestation of the importance of the temperature variation in the magnetic shape anisotropy for designing magnetic devices with (Ga,Mn)As on the sub-micrometer length scale.

\section*{Acknowledgment}
Authors thank Prof. H. Munekata and Prof. Y. Kitamoto of Tokyo Institute of Technology for kindly offering the opportunity to use their facilities for the growth of ferromagnetic semiconductors.

\end{document}